\documentclass[12pt]{article}
\pdfoutput=1
\usepackage{epsfig}
\usepackage{psfrag}
\usepackage{enumitem}
\usepackage{latexsym}
\usepackage{indentfirst}
\usepackage{fancyhdr}
\usepackage{amssymb}
\usepackage{amsmath}
\usepackage{amsfonts}
\usepackage{pifont}
\usepackage{cite}
\usepackage{bbold}
\usepackage{color}
\usepackage{graphicx}
\usepackage[center,footnotesize,hang]{subfigure}
\usepackage{url}
\usepackage{array}

\def\be{\begin{equation}}
\def\ee{\end{equation}}
\def\bc{\begin{center}}
\def\ec{\end{center}}
\def\bea{\begin{eqnarray}}
\def\eea{\end{eqnarray}}

\newcommand{\ba}{\begin{array}{c}}
\newcommand{\bad}{\begin{array}{ccc}}
\newcommand{\ea}{\end{array}}

\def\nn{\nonumber}

\begin{document}

 \begin{titlepage}
 \hfill{RM3-TH/11-13}
 \vskip 2.5cm
 \begin{center}
  {\Large\bf  Large $\theta_{13}$ from a model with broken $L_e-L_\mu-L_\tau$ symmetry}
  \end{center}
 \vskip 0.2  cm
 \vskip 0.5  cm
  \begin{center}
  {\large Davide Meloni }~\footnote{e-mail address: davide.meloni@fis.uniroma3.it}
  \\
  \vskip .2cm {\it Dipartimento di Fisica "E. Amaldi"}
  \\
  {\it Universit\'a degli Studi Roma Tre, Via della Vasca Navale 84, 00146 Roma, Italy}
  \\
  \end{center}
 \vskip 0.7cm
\begin{abstract}
\noindent
Recent data in the neutrino sector point towards a relatively large value of
the reactor angle, incompatible with a vanishing $\theta_ {13}$ at about 3$\sigma$. 
In order to explain such a result, we propose a SUSY model based on a broken $U(1)_F$ flavour symmetry
with charge $L_e-L_\mu-L_\tau$ for lepton doublets and arbitrary charge for the right-handed $SU(2)$ singlet fields,
where large deviations from the symmetric limit $\theta_{12} = \pi/4$, $\tan \theta_{23} \sim {\cal O}(1) $ and $\theta_{13} = 0$
mainly come from the charged lepton sector. We show that a description of all neutrino data is possible 
if the charged lepton mass matrix has a special pattern of complex matrix elements.
 
\end{abstract}
\end{titlepage}
%
%

\section{Introduction}
Hints for not so small $\theta_{13}$ in the neutrino sector have been recently confirmed by 
analyses of global neutrino data,
including the latest T2K \cite{T2K} and MINOS \cite{minos} results, providing $\sin \theta_{13}>0$ at 3$\sigma$ CL
\cite{Fogli:2011qn,Schwetz:2011zk}. This has triggered a lot of attention in the community,  aimed at explaining 
how such relatively large values can naturally be obtained in a consistent model for fermion masses and mixings.
Attempts in this direction have been done, for instance, in the context of GUT theories \cite{GUT}, where corrections to the 
bimaximal or tri-bimaximal mixing were responsible for large deviation of $\theta_{13}$ from zero,
or using some neutrino mass textures giving non-vanishing leading order reactor angle \cite{others}.
In this paper we want to analyze the possibility to get sizable $\theta_{13}$ from broken $U(1)_F$ flavour 
symmetry \cite{Froggatt:1978nt} with charge $L_e-L_\mu-L_\tau$ for lepton doublets \cite{Petcov:1982ya} and arbitrary right-handed charges 
\cite{altarelli}.
It is well known that, in the limit of exact symmetry, neutrino masses and mixings are
described by the following mass matrix:
\bea m_\nu =m_0
\left(
\begin{array}{ccc}
0& 1& x\\ 1& 0&0\\ x& 0&0
\end{array}
\right)\,,
\label{mass}
\eea
leading to a spectrum of inverted type, $\theta_{12} = \pi/4$, $\tan \theta_{23} = x$ (i.e. large atmospheric mixing for $x\sim {\cal O}(1)$) and $\theta_{13} = 0$.
A main problem of such texture is that two eigenvalues have the same absolute values, thus preventing the description of the 
solar mass difference. In this context, successful tentatives to solve such problems have been presented in \cite{solutions} where,
however, either the solar angle was too large compared to the current best fit point or the reactor angle was (almost) vanishing.
Corrections of order of the Cabibbo angle $\lambda$ from the charged lepton sector \cite{leptsect,Altarelli:2004jb} can simultaneously affect
the deviation of $\theta_{12}$ from maximal mixing and of $\theta_{13}$ from zero, thus allowing a sizable reactor angle,
but the question of how to naturally explain the smallness of the solar-to-atmospheric mass ratio $r$ 
(without fine-tuning among the model parameters) remained unanswered. 
In this paper we want to contribute to the discussion presenting a model where the previous issues are solved/mitigated.
In particular, we work in a SUSY framework where the  $U(1)_F$ flavour symmetry is broken by the vevs of two 
complex fields $\phi$ and $\theta$ of charge $Q_\phi=1$ and $Q_\theta=-1/2$. We show in Sect.\ref{description} that an appropriate breaking of the $L_e-L_\mu-L_\tau$
symmetry in the neutrino sector is enough to guarantee a value of $r\sim \lambda^2$ (and to preserve the leading order (LO)
prediction of large $\theta_{23}$) but not to generate the necessary
deviations for the solar and reactor angles. The corrections from the charged lepton sector are
discussed in Sect.\ref{chlepsectr} and their impact on the determination of the leptonic mixing angles 
is investigated in Sect.\ref{upmns}, where we show that complex Yukawa couplings are 
needed to make the model in agreement with the experimental data.
In Sect.5 we draw our conclusions.

\section{The neutrino sector}
\label{description}
The three families of $SU(2)$ lepton doublets $l_i$ are charged under the non-standard lepton
number $L_e - L_\mu - L_\tau$, with values:
\bea
l_i\sim L_e-L_\mu-L_\tau \sim (1,-1,-1)\nonumber\,.
\eea
From the dimension 5 operators and in the limit
of unbroken  $L_e - L_\mu - L_\tau$ symmetry, the LO lagrangian 
can be written in the (simplified) form:
\bea
{\cal L}^{LO} = \frac{1}{M}\left(x_{12} l_e l_\mu + x_{13} l_e l_\tau\right) H_u H_u + h.c.\,,
\eea
where $M$ is a large mass scale, $H_u$ represents the (uncharged) Higgs field and $x_{ij}$ are generic ${\cal O}(1)$ coefficients, that we take 
as real from now on.
The previous lagrangian gives rise to the neutrino mass matrix in eq.(\ref{mass}),
where $m_0=v^2 x_{12}/2 M$ and $x=x_{13}/x_{12}$.  
It is well know that such a  matrix has two degenerate eigenvalues $m_1 = -m_2 =m_0\, \sqrt{1+x^2}$ and a vanishing
one $m_3 = 0$; the diagonalizing matrix can be cast in the following form:
\bea U_\nu =
\left(
\begin{array}{ccc}
-\frac{1}{\sqrt{2}}&\frac{1}{\sqrt{2}} & 0 \\ \frac{1}{\sqrt{2}\sqrt{1+x^2}} & \frac{1}{\sqrt{2}\sqrt{1+x^2}} &-\frac{x}{\sqrt{1+x^2}}\\  
\frac{x}{\sqrt{2}\sqrt{1+x^2}} &\frac{x}{\sqrt{2}\sqrt{1+x^2}} &\frac{1}{\sqrt{1+x^2}}
\end{array}
\right)\,,
\label{rot}
\eea
from which 
$\tan \theta_{12} = 1$, $\tan \theta_{23} = x$ and $\sin \theta_{13} = 0$.
From the expressions of the atmospheric mass difference $\Delta m^2_{atm}=|m_1|^2-|m_3|^2$, we easily get:
\bea
x^2 = \frac{\Delta m^2_{atm}}{m_0^2} - 1 \,,
\eea
so that, for a natural value $x \sim {\cal O}(1)$ (needed also to fit the
atmospheric angle) we can fix the overall scale $m_0 = 0.035$ eV.

In order to split the solar mass difference and reproduce a correct value of $r$, we need to fill 
some of the vanishing matrix elements of eq.(\ref{mass}) with entries of appropriate order of magnitude. 
To this aim, it would be enough to assume that the $U(1)_F$ symmetry is broken by the vev of one scalar field;
however, as it will be discussed at length in Sect.\ref{chlepsectr}, we also want to generate 
a large $\theta_{13}$ (and avoid large contributions to $\mu \to e \gamma$ fro the charged lepton sector); this forces 
to assume that the $U(1)_F$ symmetry is broken by the vev of {\it two} fields $\phi$ and $\theta$ of charge 
$Q_\phi=1$ and $Q_\theta=-1/2$. 
To allow for non-vanishing vevs, we assume to work in the limit of exact SUSY; in this case,
the vev of the scalar fields are determined by setting to zero the corresponding scalar potential \cite{Altarelli:2008bg}.
For the flavon fields $\theta$ and $\phi$, related to the Froggatt-Nielsen symmetry, the non vanishing
vevs are determined by the D-term associated with the $U(1)_{FN}$ symmetry:
$V_D=\frac{1}{2}(M_{FI}^2- g_{F}\vert\theta\vert^2-g_{F}\vert\phi\vert^2)^2$,
where $g_{F}$ is the gauge coupling constant of $U(1)_{F}$ and $M_{FI}^2$ denotes the contribution of the Fayet-Iliopoulos term; the condition
for having SUSY minima, $V_D=0$, gives the relation for non-vanishing vev of the
combination $M_{FI}^2= g_{F}\vert\theta\vert^2+g_{F}\vert\phi\vert^2$. It is then natural to assume a common order 
of magnitude for both vevs, $\langle \Phi \rangle /\Lambda\equiv \lambda$, $\lambda$ being the Cabibbo angle.
With this in mind, the next-to-leading order (NLO) contributions 
to the neutrino mass matrix come from:
\bea
{\cal L}^{NLO} = \frac{1}{M}\left(x_{2}^\prime l_\mu l_\mu\frac{\phi^2}{\Lambda^2} + x_{3}^\prime l_\mu l_\tau\frac{\phi^2}{\Lambda^2}
+ x_{4}^\prime l_\tau l_\tau\frac{\phi^2}{\Lambda^2}  + x_{1}^\prime l_e l_e\frac{\theta^4}{\Lambda^4}\right) H_u H_u + h.c.
\eea
so that
\bea m_\nu =m_0\,
\left(
\begin{array}{ccc}
x_1 \, \lambda^4& 1& x\\ 
1&x_2 \, \lambda^2&x_3 \, \lambda^2\\ 
x& x_3 \, \lambda^2&x_4 \, \lambda^2
\end{array}
\right)\,,
\label{invh1}
\eea
with $x_i=x_i^\prime/x_{12}$.
We see that the charge of $\phi$ is needed to fill the 2-3 sub-block with entries of ${\cal O}(\lambda^2)$, whereas at the same order the contribution
of the $\theta$ field is negligible. 
%
From standard perturbation theory, we get the following eigenvalues (in units of $m_0$):
\bea
m_1 &=&\sqrt{1+x^2} + \left[\frac{x_2 +x (2  x_3 + x x_4)}{2
   \left(1+x^2\right)}\right]\,\lambda^2  \nn \\
m_2 &=&-\sqrt{1+x^2} + \left[\frac{x_2 +x (2  x_3 + x x_4)}{2
   \left(1+x^2\right)}\right]\,\lambda^2   \\
m_3 &=&\left[\frac{x^2 x_2-2 x x_3 +x_4}{1+x^2}\right]\,\lambda^2\nn \,.
\eea
Notice that the corrections to $m_{1,2}$ are exactly the same. We can compute the small parameter 
$r=\Delta m^2_{sol}/\Delta m^2_{atm}$ in the limit of real parameters, obtaining:
\bea
r= -2 \left[\frac{2 x x_3 +x_2+x^2 x_4}{
   \left(1+x^2\right)^{3/2}}\right]\,\lambda^2\,.
\eea
For ${\cal O}(1)$ parameters,  a natural suppression of $r$ by $\lambda^2$ is achieved and the usual problem of having $r\sim {\cal O}(1)$ (especially when
using non-abelian groups with neutrinos in triplet representations) is circumvented here.
Notice that the inverted mass ordering predicted by the model can be already tested after the second phase 
of the T2K experiment \cite{Huber:2002rs} or, possibly, at future Neutrino Factories or $\beta$ beams \cite{NF:2011aa}.

The normalized eigenvectors corresponding to $m_1$ and $m_2$ receive ${\cal O}(\lambda^2)$ corrections whereas, at the same order, only the first 
component of the third eigenvector is modified compared to the unbroken limit. It it easy to extract the values 
of the mixing angles (before the diagonalization of the charged leptons):
\bea
\sin \theta^\nu_{13}&=&  \left[\frac{x (x x_3+x_2-x_4)-x_3}{\left(1+x^2\right)^{3/2}}\right]\,\lambda^2 \nn \\
\tan\theta^\nu_{12}&=& 1+ \left[\frac{x (x x_4+2 x_3)+x_2}{2
   \left(1+x^2\right)^{3/2}}\right]\,\lambda^2 \sim  1 - \frac{r}{4}\label{resneu} \\
\tan \theta^\nu_{23}&=& x  \nn\,.
\eea
We see that,  
although the atmospheric angle is still of  ${\cal O}(1)$, the solar angle remains too large (and always smaller than maximal mixing, due to the condition $r>0$) 
and that the reactor 
angle is as small as $\lambda^2$. The corrections from the charged lepton sector must be large, of ${\cal O}(\lambda)$, to shift 
both $\theta_{12}$ and  $\theta_{13}$ closer to their recent best fit values \cite{Fogli:2011qn} while not destroying 
a large $\theta_{23}$. Notice that the rotation from such a sector also affects the radiative decays of muons and taus. In particular,
the correction in the $(1-2)$ sector is the most relevant one and should be taken under control to avoid a large contribution
to $\mu \to e \gamma$.
In fact, it has been shown in \cite{Feruglio:2008ht} and \cite{Altarelli:2009gn}  that, in a effective theory approach, 
the amplitudes for the radiative decays of the charged leptons 
depend on the off-diagonal elements of the dipole operator ${\cal M}$,
in a basis where the mass matrix of the charged leptons is diagonal.
In the case of $\mu \to e \gamma$, the relevant matrix element is ${\cal M}_{12}$, suppressed 
by a cut-off scale and directly related to the $(1-2)$ charged lepton mixing. In SUSY theories, 
if we want the cut-off scale at the level 
of $\sim$ 1 TeV, the $(1-2)$ mixing should be at least of ${\cal O}(\lambda^2)$
to avoid a large ${\cal M}_{12}$ and then a large $\mu \to e \gamma$ branching ratio.
 This means that the deviations from $\theta_{12} = \pi/4$ can only originate from a different subsector
of the charged lepton mass matrix. 

\section{The charged lepton sector}
\label{chlepsectr}
We start with a general consideration about the charged lepton rotation \cite{Altarelli:2004jb}.
Let us define a generic $3\times3$ mixing matrix in terms of rotations in three different subspaces:
\bea
\label{genericU}
\tilde U~=~ 
\left(
\begin{matrix}
1&0&0 \cr 0&c_{23}^e&s_{23}^e\cr0&-s_{23}^e&c_{23}^e 
\end{matrix}
\right)
\left(\begin{matrix}
c_{13}^e&0&s_{13}^e e^{i\delta} \cr 0&1&0\cr -s_{13}^e e^{-i\delta}&0&c_{13}^e     
\end{matrix}
\right)
\left(\begin{matrix}c_{12}^e&s_{12}^e&0 \cr -s_{12}^e&c_{12}^e&0\cr 0&0&1   \end{matrix}
\right)\,.
\label{ufi}
\eea
where we used the superscript $e$ to identify the angles in the charged lepton sector. In the left-right basis, 
the mass matrix is diagonalized by a bi-unitary transformation:
\bea
m_\ell = U_L \, m_\ell ^D \, U_R^\dagger\,.
\eea
Using for $U_L$ the form in eq.(\ref{genericU}) and disregarding possible CP phases, we obtain:
\bea
m_\ell = \left(
\begin{matrix}
 c_{12}^e c_{13}^e m_e & s_{12}^e c_{13}^e m_\mu & s_{13}^e m_\tau \cr
m_e (-c_{23}^e s_{12}^e - c_{12}^e s_{13}^e s_{23}^e) & m_\mu (c_{12}^e c_{23}^e - s_{12}^e s_{13}^e s_{23}^e) & 
  c_{13}^e s_{23}^e m_\tau \cr
m_e (-c_{12}^e c_{23}^e s_{13}^e + s_{12}^e s_{23}^e) & m_\mu (-c_{23}^e s_{12}^e s_{13}^e - c_{12}^e s_{23}^e) &
  c_{13}^e c_{23}^e m_\tau
\end{matrix}
\right)\, U_R\,.
\eea
The right-handed rotation is not involved in the neutrino mixing; inspired by the minimal SU(5) relation $m_e=m_d^T$ and by the fact that 
the CKM matrix is almost 
diagonal, we tentatively take $U_R \sim 1$.
Now we want to introduce in $m_\ell$ appropriate expressions for the masses and angles.
According to the previous discussion, the needed pattern of corrections are guarantee if 
we take $s_{12}$ to be smaller than ${\cal O}(\lambda^2)$ and $s_{13}\sim {\cal O}(\lambda)$ \cite{Altarelli:2004jb}; 
for the moment, we do not specify the value of the atmospheric angle. Finally, 
we factorize out the $\tau$ mass from $m_\ell$ and use the mass ratios
\bea
\nn
m_e : m_\mu : m_\tau = \lambda^5 : \lambda^2 : 1\,
\eea
to get:
\bea
\label{emmee}
m_\ell \sim m_\tau\, \left(
\begin{matrix}
 \lambda^5 & \lambda^{\lesssim 4} & \lambda \cr
\lambda^6 & c_{23}^e\lambda^2 &  s_{23}^e \cr
\lambda^6 & -s_{23}^e\lambda^2&c_{23}^e 
\end{matrix}
\right)\,.
\eea
By construction, this matrix provides a left-handed rotation of type (taking only the leading contributions in $\lambda$):
\bea
\label{dian}
U_L \sim \left(
\begin{matrix}
 1 & \lambda^{\lesssim 2} & \lambda \cr
-s_{23}^e\lambda & c_{23}^e  &  s_{23}^e \cr
-c_{23}^e\lambda  & -s_{23}^e &c_{23}^e 
\end{matrix}
\right)\,,
\eea
which generates the wanted corrections to the solar and reactor angles while avoiding large contributions to $\mu \to e \gamma$. 
The peculiarity of such a matrix comes from the (2-3) sub-block, whose matrix elements are strongly correlated in terms of relative signs and 
magnitudes: in fact,  there is only one independent element, dictated by the $\theta_{23}^e$ angle.
If we were able to naturally obtain eq.(\ref{emmee}), then the neutrino mixing angles, from $U_{PMNS}=U_L^\dagger\,U_\nu$, 
would have the following expressions (disregarding ${\cal O}(\lambda^2)$ corrections
stemming from neutrino NLO corrections):
\bea
\label{belle1}
\sin \theta_{13} &\sim&\left| \frac{-c_{23}^e + s_{23}^e x}{\sqrt{1+x^2}}\right|\,\lambda \nn \\
\tan \theta_{12}&\sim& 1 +  2 \left(\frac{s_{23}^e + c_{23}^e x}{\sqrt{1+x^2}}\right)\,\lambda \\
\tan \theta_{23} &\sim& \left|\frac{s_{23}^e + c_{23}^e x}{-c_{23}^e + s_{23}^e x }\right|  \nn\,.
\eea
For $\theta_{23}^e \to 0 $   (and still $x\sim {\cal}O(1)$), we would get the good results:
 \bea
\label{belle}
\sin \theta_{13} &\sim&\frac{\lambda}{\sqrt{2}} \qquad 
\tan \theta_{12}\sim 1 + {\cal O} \left(\sqrt{2}\, \lambda\right) \qquad
\tan \theta_{23}  \sim  1\,.
\eea
However, the matrix in eq.(\ref{emmee}) is difficult to implement in models based on $U(1)_F$, which only allows to accommodate  the order of magnitude of the matrix elements. 
Moreover, as it will be clear later, in models with broken $L_e -L_\mu - L_\tau$ symmetry, the charges of the mass matrix elements $(m_\ell)_{23}$ and 
$(m_\ell)_{33}$ are exactly the same and we cannot generate any hierarchy among them without invoking additional ingredients.
In the following, we then look for charge assignments of the right-handed fields which allow to reproduce a structure as similar as possible to 
eq.(\ref{emmee}) in terms of powers of $\lambda$; this is as:
\bea
\label{emmee2}
m_\ell \sim \left(
\begin{matrix}
 \lambda^5 & \lambda^{\lesssim 4} & \lambda \cr
0 &  \lambda^2 & 1 \cr
0 &  \lambda^2&1
\end{matrix}
\right) \,.
\eea
We emphasize that a matrix like eq.(\ref{emmee2}) {\it does not} naturally produce mixing angles in agreement with eq.(\ref{belle1}).
The operators contributing to the mass matrix are of the form:
\bea
\label{oper}
(m_\ell)_{ij}= a_{ij}\, l_i\,l_j^c\, \left(\frac{\langle \phi \rangle}{\Lambda}\right)^{\alpha_{ij}}\,\left(\frac{\langle \theta \rangle}{\Lambda}\right)^{\beta_{ij}}\,H_d
\sim a_{ij}\, l_i\,l_j^c\;\lambda^{\alpha_{ij}+\beta_{ij}}\, ,
\eea
where $\alpha_{ij}$ and $\beta_{ij}$ are positive integers controlling the suppression of each matrix element in terms of $\lambda$, 
with $\alpha_{ij}+\beta_{ij}$ being the minimum allowed sum (otherwise we would take into account higher order combinations of
flavon fields) and $a_{ij}$ are complex ${\cal O}(1)$ coefficients. $H_d$ is taken to be uncharged.
With the $U(1)_F$ assignment for the leptons as:
\bea
l_i\sim L_e-L_\mu-L_\tau \sim (1,-1,-1)\nonumber \\
l^c \sim (Q_e, Q_\mu, Q_\tau) \label{assign} \,,
\eea
and reminding that $Q_\phi=+1$ and $Q_\theta=-1/2$,
the charges of the operators in eq.(\ref{oper}) are as follows:
\bea
Q_{m_\ell}= \left(
\begin{array}{ccc}
1+Q_e+\alpha_{11}  - \beta_{11} /2& 1+Q_\mu+\alpha_{12}  - \beta_{12}/2 & 
1+Q_\tau+\alpha_{13}  - \beta_{13}/2 \\
-1+Q_e+\alpha_{21}  - \beta_{21}/2 & -1+Q_\mu+\alpha_{22}  - \beta_{22}/2 & 
-1+Q_\tau+\alpha_{23}  - \beta_{23}/2  \\
-1+Q_e+\alpha_{21}  - \beta_{21}/2 & -1+Q_\mu+\alpha_{22}  - \beta_{22}/2 & 
-1+Q_\tau+\alpha_{23}  - \beta_{23}/2 \\
\end{array}
\right)\,.\label{melle2}
\eea
Notice that, having the $\mu$ and $\tau$ the same $L_e-L_\mu-L_\tau$ charge, the operators contributing to the second and third row
of the mass matrix must have exactly the same operator structure in terms of flavon fields or, in other words, the coefficients $\alpha$ and $\beta$
must be equal.

The choice $Q_\tau=1$ is untenable; in fact, in this case $(Q_{m_\ell})_{33}$ is already invariant (and then
$\alpha_{23}  + \beta_{23} = 0$) whereas $(Q_{m_\ell})_{13} = 2 +\alpha_{13}  - \beta_{13}/2$
and the minimum choice of charges to get $(Q_{m_\ell})_{13} =0$ is $(\alpha_{13},\beta_{13})=(0,4)$;
in the mass matrix, this would mean that $(m_\ell)_{13}$ would  be suppressed by $\lambda^4$ compared to  $(m_\ell)_{33}$,
thus destroying
the wanted  hierarchy as in eq.(\ref{emmee2}). One possible solution 
is to take $Q_\tau=0$; in this case $(\alpha_{23},\beta_{23})=(1,0)$ and $(\alpha_{13},\beta_{13})=(0,2)$
and the last column of eq.(\ref{emmee2}) is 
\bea
\left(
\begin{array}{c}
\left(\langle \theta \rangle/\Lambda\right)^2 \\
\langle \phi \rangle/\Lambda \\
\langle \phi \rangle/\Lambda\\
\end{array}
\right)\sim \lambda
\left(
\begin{array}{c}
\lambda \\
1 \\
1\\
\end{array}
\right)\,.
\eea
With the same reasoning, we can fix the other charges $Q_e$ and $Q_\mu$ to reproduce eq.(\ref{emmee2}) and then the hierarchy among the charged lepton masses.
It is easy to show that the choice $(Q_e, Q_\mu)=(-7,5/2)$ gives the correct result:
\bea
\label{emmee3}
m_\ell \sim \lambda\left(
\begin{matrix}
 \langle \phi \rangle^5/\Lambda^5 & \langle \theta \rangle^6/\Lambda^6 & \langle \theta \rangle/\Lambda\\
\langle \phi \rangle^7/\Lambda^7 & \langle \theta \rangle^2/\Lambda^2 & 1 \\
 \langle \phi \rangle^7/\Lambda^7 & \langle \theta \rangle^2/\Lambda^2 & 1
\end{matrix}
\right)  = \lambda\left(
\begin{matrix}
 \lambda^5 & \lambda^6 & \lambda\\
\lambda^7 & \lambda^2 & 1 \\
\lambda^7 & \lambda^2 & 1
\end{matrix}
\right)\,.
\eea
The set of charges that can reproduce the observed charged lepton masses and a structure similar to eq.(\ref{emmee}) are summarized in Tab.(\ref{charges}).
\begin{table}[h!]
\begin{center}
\begin{tabular}{|c|c|c|c|c|c|c|c|c|}
\hline
& $l_e$& $l_{\mu}$& $l_{\tau}$ & $l_e^c$& $l_{\mu}^c$& $l_{\tau}^c$ & $\phi$ & $\theta$ \\
\hline
\hline 
$U(1)_F$& 1 & -1& -1& -7&5/2 &0 &1 &-1/2\\
\hline
\end{tabular}
\end{center}
\label{default}
\caption{\it \label{charges} Set of charges that can accommodate the observed charged lepton masses and neutrino mixings.}
\end{table}

\section{The $U_{PMNS}$ mixing matrix}
\label{upmns}
\begin{figure}[h!]
\begin{center}
\includegraphics[scale=.6]{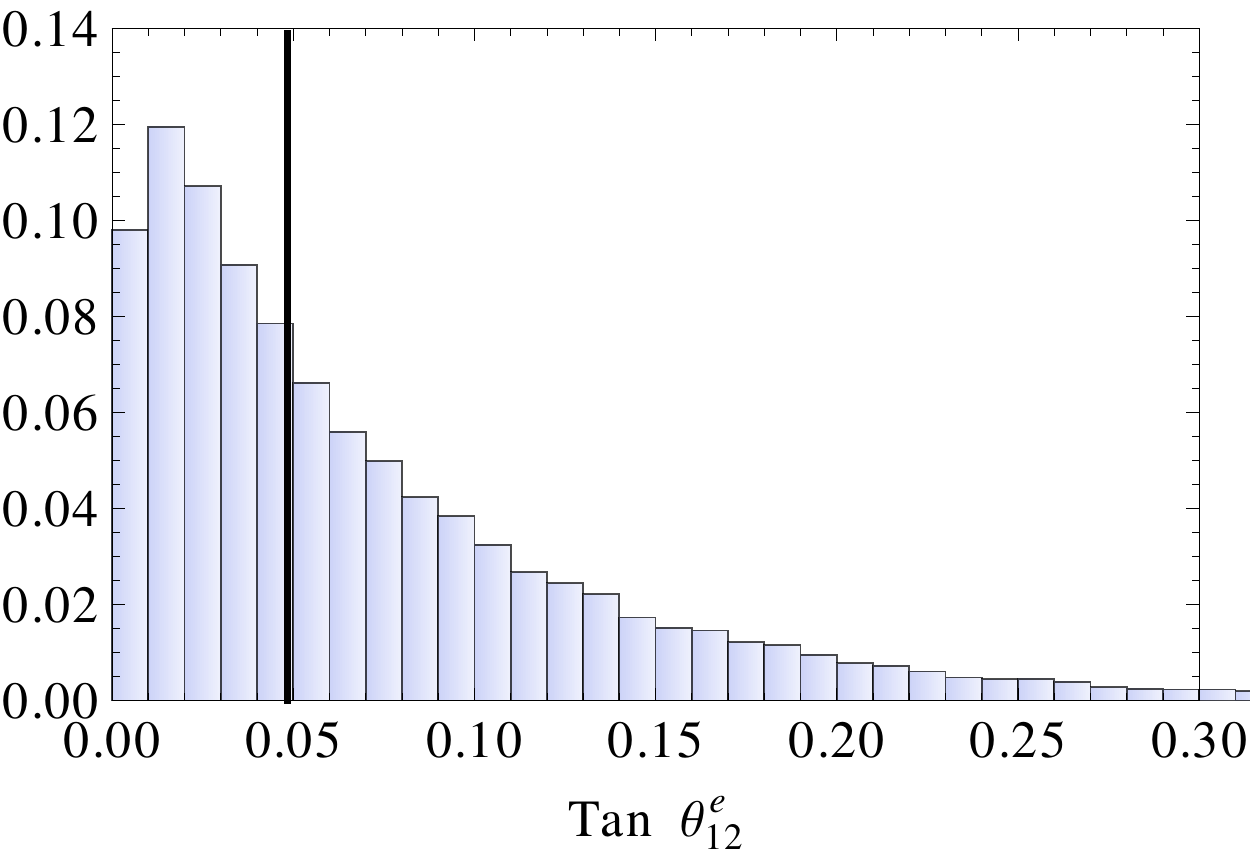} 
\caption{\label{12lept} \it Distribution for $\tan \theta_{12}^e$ (arbitrary units). The black line indicates the value  $\lambda^2$.}
\end{center}
\end{figure}

\begin{figure}[h!]
\begin{center}
\includegraphics[scale=.5]{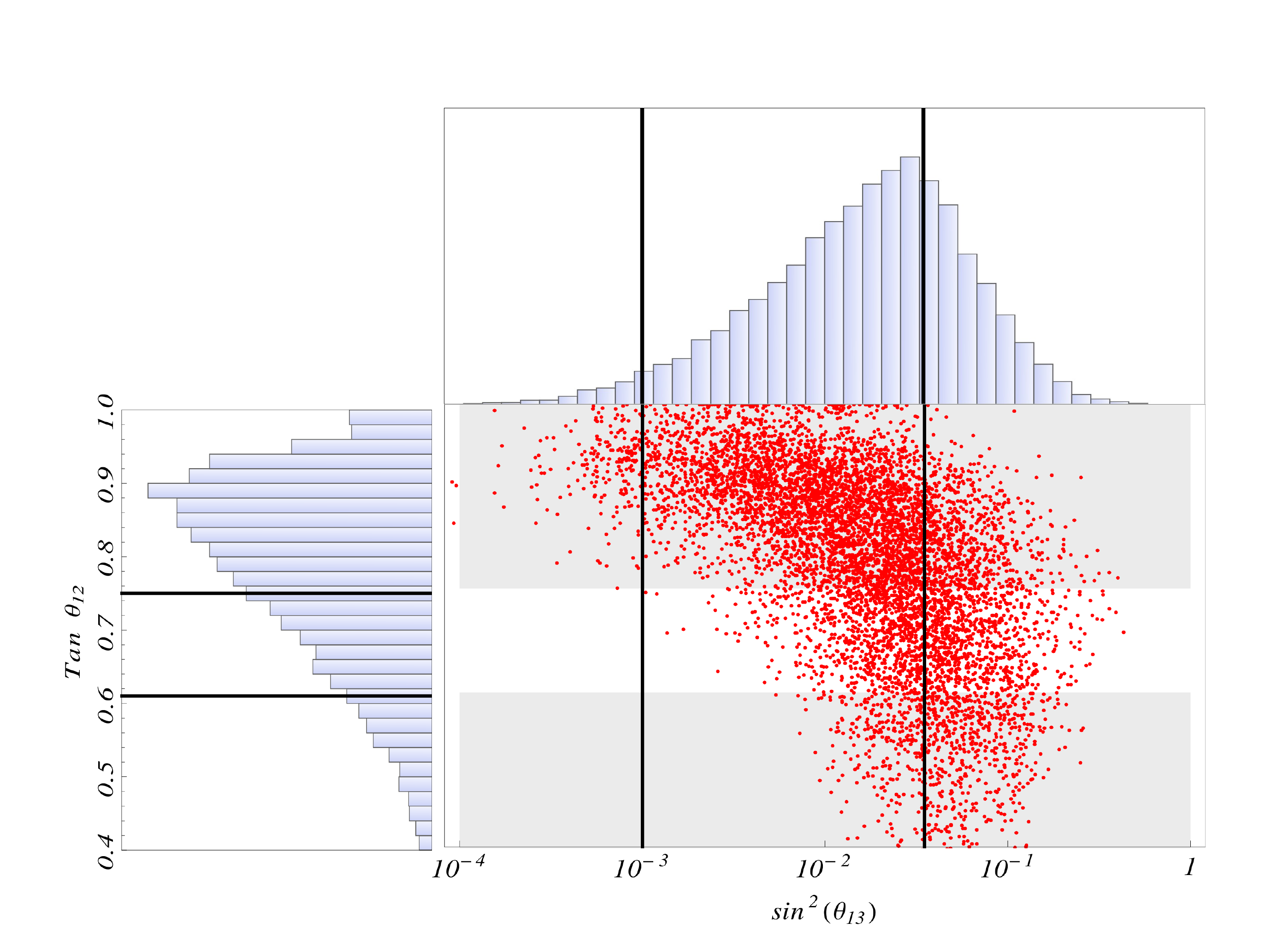} 
\caption{\label{plots} \it Central panel: scatter plot in the variables $(\sin^2\theta_{13},\tan\theta_{12})$ as obtained from our model. 
The gray bands represents the regions excluded by the experimental data on $\tan\theta_{12}$ at 3$\sigma$ \cite{Schwetz:2011zk}
whereas the vertical lines enclose the 3$\sigma$ range on $\sin^2\theta_{13}$.
See text for further details. Upper and left panels: distributions of the variables $\sin^2\theta_{13}$ and $\tan\theta_{12}$, respectively, and 
their 3$\sigma$ allowed regions, enclosed into the solid (black) lines.}
\end{center}
\end{figure}

In the language of $U(1)_F$ flavour models, the mass matrix produced by our charge assignment is as follows:
\bea
\label{emmee5}
m_\ell \sim  m_\tau \,
\left(
\begin{matrix}
a_{11} \lambda^5 & a_{12}\lambda^6 & a_{13}\lambda\\
a_{21}\lambda^7 & a_{22}\,e^{i\phi_{22}}\lambda^2 & a_{23}\,e^{i\phi_{23}} \\
a_{31} \lambda^7 & a_{32}\,e^{i\phi_{32}}\lambda^2 & 1
\end{matrix}
\right)\,,
\eea
where we have explicitely shown the phases of the $(22),(23)$ and $(32)$ matrix elements (phases into the other entries are
unimportant or even irrelevant for our reasoning); all $a_{ij}$ are generic ${\cal O}(1)$ coefficients\footnote{We note that the mass matrix $m_\ell$ implies a right-handed charged lepton rotation
very close to the identity, as assumed in Sect.3.}.
For our discussion, it is important to quote the leptonic $\theta_{12}^e$ (obtained after diagonalizing
$m_\ell m_\ell^\dagger$) and the three neutrino mixing angles, including the corrections from the charged lepton sector:
\bea
\tan \theta_{12}^e &=&  |a_{13}|  \left[\frac{a_{22}^2\, a_{23}^2+a_{32}^2+2\, a_{22}\, a_{23}\, a_{32} \,\cos (\phi_{23}-\phi_A)}
{(1+a_{23}^2) \,D}\right]^\frac{1}{2}\,\lambda \nn \\ && \nn \\
\sin \theta_{13} &=& |a_{13}|  \left[\frac{a_{22}^2 + a_{32}^2\,x^2+2\, a_{22}\, a_{32} \,x\,\cos \phi_A}
{(1+x^2) \,D}\right]^\frac{1}{2}\,\lambda 
\\ && \nn \\
\tan \theta_{12} &=&1+\frac{2 a_{13}  \left[a_{22} (a_{22} x-a_{32} (a_{23} x \cos (\phi_{23}-\phi_A)+\cos\phi_A))+
a_{23} a_{32}^2 \cos\phi_{23}\right]}{D \sqrt{1+x^2}}\,\lambda \nn \\ && \nn \\
\tan \theta_{23} &=&\sqrt{\frac{a_{23}^2+2 a_{23} x \cos\phi_{23}+x^2}{1+a_{23}^2 x^2-2 a_{23} x \cos\phi_{23}}} \, \nn ,
\eea 
where we have used the short-hand notations:
\bea
D=a_{22}^2-2 a_{22} a_{23} a_{32} \cos (\phi_{23}-\phi_A)+a_{23}^2 a_{32}^2
\qquad
\phi_A=\phi_{22}-\phi_{32}\nn\,.
\eea
In  spite of their apparent complexity, the previous relations offer a simple way 
to reconcile the model with the experimental data.  First of all, we observe that 
$\theta_{12}^e$, $\theta_{13}$ and the deviation of $\theta_{12}$ from maximal mixing are all of ${\cal O}(\lambda)$;
to achieve a suppression for the leptonic $\theta_{12}$ we need $\cos (\phi_{23}-\phi_A)\sim -1$; on the other
hand, this is not enough to guarantee an almost maximal atmospheric mixing, unless $a_{23}$ and $x$ are
both very different from  ${\cal O}(1)$; we then require the additional condition $\cos \phi_A\sim 0$.
%
With these positions, the analytical structure of the mixing angles is simple and allows a clear understanding of  
the outcome of such a procedure:
\bea
\tan \theta_{12}^e &=& \frac{|-a_{22}\, a_{23}+a_{32}|}{a_{22}+ a_{23} a_{32}}
\frac{|a_{13}|}{\sqrt{1+a_{23}^2}}\,\lambda \nn \\ && \nn \\
\sin \theta_{13} &=& \left(\frac{|a_{13}|}{a_{22}+a_{23}\,a_{32}}\right) 
\sqrt{\frac{a_{22}^2+a_{32}^2 \,x^2}{1+x^2}}
\,\lambda \\ && \nn \\
\tan \theta_{12} &=&1+\frac{2\,a_{13}\,a_{22}\,x}{(a_{22}+a_{23}\,a_{32})\,\sqrt{1+x^2}}  \,\lambda \nn \\ && \nn \\
\tan \theta_{23} &=&\sqrt{\frac{a_{23}^2+  x^2}{1+a_{23}^2\,x^2}} \, \nn .
\eea 
Bearing in mind that the parameters $a_{22,23,32}$ are all positive, we see that a cancellation in the numerator of $\tan \theta_{12}^e$
occurs; moreover, we expect: 
\bea
\label{pred}
\sin \theta_{13} \sim \frac{\lambda}{2} \qquad \tan \theta_{12}\sim 1\pm \frac{\lambda}{\sqrt{2}}\qquad \tan \theta_{23} \sim 1\,,
\eea
and then a clear correlation among the solar and reactor angles:
\bea
\label{corr}
\theta_{12}\sim \frac{\pi}{4} \pm \frac{\theta_{13}}{\sqrt{2}}\,.
\eea
It is interesting to observe that, in the limit of real Yukawa parameters, it would be very difficult 
to simultaneously take under control both $\theta_{12}^e$ and $\theta_{23}$, unless strong cancellations among the 
relevant $a_{ij}$ are invoked.
Our analytical results are confirmed by a numerical evaluation of the mixing angles, obtained extracting randomly the values of the parameters 
$a_{ij}$ and $x,x_i$ of the mass matrices (\ref{emmee5}) and (\ref{invh1}), respectively,
in 
the interval $[1/3,3]$ and the phases $(\phi_{23},\phi_A)$ around $(3\pi/2,\pi/2)\pm 0.1$; we do not impose any other constraints.
In Fig.(\ref{12lept}) we show the distributions of values for $\tan \theta_{12}^e$; we see that many points fall in the region 
below $\lambda^2$ (indicated with a solid vertical line), corresponding to almost 35\% of the total number 
of realizations. 
\noindent\\
In Fig.(\ref{plots}) we present the main result of this paper, namely the correlation among $\theta_{12}$ and $\theta_{13}$ in the 
$(\sin^2\theta_{13},\tan\theta_{12})$ plane\footnote{We only show
values of $\tan \theta_{12}\le 1$; an almost symmetric region is populated at $\tan \theta_{12}>1$.}. In the central plot 
we show the simultaneous determinations of both angles; vertical solid (black) lines enclose the 3$\sigma$ bounds on $\theta_{13}$ from 
\cite{Schwetz:2011zk} whereas the gray horizontal bands are the regions excluded by the experimental data on $\theta_{12}$ 
(also at 3$\sigma$ from \cite{Schwetz:2011zk}), that is 
\bea
0.001\le &\sin^2 \theta_{13}&\le 0.035 \nn\\  \\
0.61\le &\tan \theta_{12}&\le 0.75 \nn\,.
\eea
We clearly see that the correlation shown in eq.(\ref{corr}) is in fact realized in our model: large values of the reactor angle bring 
the solar one into the experimentally allowed region whereas, for small $\theta_{13}$, $\theta_{12}$ is still close to maximal mixing.
In the upper and left panels we also display the distributions of the related variables $\sin^2\theta_{13}$ and $\tan \theta_{12}$, respectively,
with the vertical solid (black) lines enclosing their 3$\sigma$ bounds. It is interesting to observe that $\sim$67\% of the 
points in the  $\sin^2\theta_{13}$ distribution falls in the current allowed range whereas such a number decreases to  $\sim$30\%
for $\tan \theta_{12}$.
We also notice that the distributions of the angles have the picks in the region around the values indicated in eq.(\ref{pred}).
For the atmospheric angle we found no significant correlations with the other mixing parameters and a distribution of values 
with more than 50\% in the allowed 3$\sigma$ range. We do not show the corresponding plots.

Finally, we analyze the model predition for the 
effective mass $m_{ee}$ appearing in the neutrinoless double $\beta$ decay amplitude;
since the lightest neutrino mass $m_3$ is at most of ${\cal O}(\lambda^2)$ (and $s_{13}\sim {\cal O}(\lambda)$), we can safely neglect its contribution to the 
effective mass $m_{ee}$ appearing in the neutrinoless double $\beta$ decay amplitude, which then reads:
\bea
|m_{ee}|\sim \sqrt{\Delta m^2_{atm}}\,(c_{12}^2+s_{12}^2\,e^{2 i \alpha})\,,
\eea
where $\alpha$ is the Majorana phase associated to $m_2$. Since we choose to work with 
real parameters in the neutrino mass matrix, $\alpha$ is generated by the charged lepton rotation.
In a basis where they are diagonal, the condition $(\phi_{23},\phi_A) = (3\pi/2,\pi/2)$ gives 
$\alpha=0$; however, like in the case for the Dirac CP phase, when we perturb the previous conditions
any value of $\alpha$ in $[0,2 \pi]$ can be generated and we get:
\bea
|m_{ee}|\sim (1.9 - 5.1) \times 10^{-2} \,{\text eV}\,.
\eea 

\section{Conclusions}
In this paper we presented a model for lepton masses and mixings based on a $U(1)_F$ flavour symmetry with charge $L_e-L_\mu-L_\tau$.
The symmetry is broken by the vevs of two scalar fields and, thanks to the corrections from the charged lepton sector, 
we get a large $\theta_{13}\sim {\cal O}(\lambda)$, while the other mixing
angles can be made compatible with the experimental data. For this to happen, we used the freedom given by the complex phases of the Yukawa 
couplings defining the 
charged lepton mass matrix; we observed that, under certain hypothesis, we are able to maintain the atmospheric angle close to 
maximal mixing and to lower the value of the leptonic 12 rotation at least at  ${\cal O}(\lambda^2)$,
necessary to avoid dangerous contributions to the rare decays of the charged leptons.
We also obtained a naturally small value of the solar-to-atmospheric mass difference ratio $r$,
at the level of $\lambda^2$. 
\section{Acknowledgments}
We thank Guido Altarelli for some interesting comments and discussions. 
We also acknowledge  MIUR (Italy) for financial support
under the program "Futuro in Ricerca 2010 (RBFR10O36O)", 
and the CERN Theory Division where this work has been completed. 

\end{document}